\documentclass[aps,prfluids,amssymb,amsmath,onecolumn,nobibnotes]{revtex4-2}

\usepackage{float}
\usepackage{booktabs}
\usepackage{graphicx}
\usepackage{setspace}
\usepackage{epstopdf}
\usepackage{bm}
\usepackage{textcomp}
\usepackage{natbib}
\usepackage{xcolor}
\usepackage{longtable}

\newcommand{\midgrad}{\left.-\partial_z\overline{T}\right|_{z=0.5}}

\begin{document}

\title{Laboratory exploration of heat transfer regimes in rapidly rotating turbulent convection}

\author{Jonathan S.~Cheng}
\altaffiliation{Current affiliation: Department of Mechanical Engineering, University of Rochester, Rochester, NY 14627, USA}
\affiliation{Fluids and Flows group, Department of Applied Physics and J.M. Burgers Center for Fluid Dynamics, Eindhoven University of Technology, P.O. Box 513, 5600 MB Eindhoven, Netherlands}
\author{Matteo Madonia}
\affiliation{Fluids and Flows group, Department of Applied Physics and J.M. Burgers Center for Fluid Dynamics, Eindhoven University of Technology, P.O. Box 513, 5600 MB Eindhoven, Netherlands}
\author{Andr\'es J.~Aguirre Guzm\'an}
\affiliation{Fluids and Flows group, Department of Applied Physics and J.M. Burgers Center for Fluid Dynamics, Eindhoven University of Technology, P.O. Box 513, 5600 MB Eindhoven, Netherlands}
\author{Rudie P.~J.~Kunnen}
\email{r.p.j.kunnen@tue.nl}
\affiliation{Fluids and Flows group, Department of Applied Physics and J.M. Burgers Center for Fluid Dynamics, Eindhoven University of Technology, P.O. Box 513, 5600 MB Eindhoven, Netherlands}

\date{\today}

\begin{abstract}
We report heat transfer and temperature profile measurements in laboratory experiments of rapidly rotating convection in water under intense thermal forcing (Rayleigh number $Ra$ as high as $\sim 10^{13}$) and unprecedentedly strong rotational influence (Ekman numbers $E$ as low as $10^{-8}$). Measurements of the mid-height vertical temperature gradient connect quantitatively to predictions from numerical models of asymptotically rapidly rotating convection, separating various flow phenomenologies. Past the limit of validity of the asymptotically-reduced models, we find novel behaviors in a regime we refer to as rotationally-influenced turbulence, where rotation is important but not as dominant as in the known geostrophic turbulence regime. The temperature gradients collapse to a Rayleigh-number scaling as $Ra^{-0.2}$ in this new regime. It is bounded from above by a critical convective Rossby number $Ro^*=0.06$ independent of domain aspect ratio $\Gamma$, clearly distinguishing it from well-studied rotation-affected convection.
\end{abstract}

\maketitle

\section{Introduction}
Convectively driven, rotationally constrained flows are the foundation of many geophysical and astrophysical processes, from dynamo action in Earth's molten iron core \cite{Glatzmaier95} to atmospheric patterns in gas giants \cite{Heimpel05}. These systems are massive, complex, and remote from most direct measurements, meaning that our understanding of their flows depends on greatly simplified models. Perhaps the most fundamental of such models is the canonical problem of rotating Rayleigh-B\'{e}nard convection, where a layer of fluid is subject to an unstable vertical temperature gradient and rotated about a vertical axis. Even in this reduced problem, though, vastly different flows emerge depending on the relative strength of rotational and convective forces, and their properties must be understood before performing any kind of geophysical extrapolation. While rotating convection is well-studied at moderate degrees of thermal forcing and rotation in laboratory experiments \cite{Rossby69,Liu97,Kunnen08,Zhong09} and direct numerical simulations (DNS) \cite{Julien96,Schmitz09}, a substantial parameter gap separates such studies from the extreme conditions in planets \cite{Schubert11,RobertsKing13}.
 
Recent studies have aimed to bridge this gap by employing large-scale experimental setups \cite{Weiss11b,King12,Ecke14,Cheng15} and high resolution simulations \cite{King12,Favier14,Stellmach14,Kunnen16}. Though gains may appear marginal in the planetary context, these studies have, in fact, manifested a plethora of novel behaviors absent from earlier studies. This is particularly true of regimes where both thermal forcing and rotation (described respectively by the Rayleigh number, $Ra$, and Ekman number, $E$, defined below) strongly affect the flow but neither dominates. Estimates of the governing parameters in planetary fluid layers indicate that understanding these regimes may be the key to solidifying the relationship between rotating convection models and geophysical systems \cite{Aurnou15,Cheng15}. Achieving strong enough rotational influence to establish these regimes, however, is no easy task: to date, our understanding of them relies primarily on simulations of the asymptotically-reduced equations, a set of equations rescaled in the limit of infinitely rapid rotation \cite{Sprague06}.

In this study, we analyze 70 new nonrotating and rotating convection experimental data points from the TROCONVEX laboratory setup \cite{Cheng18}. Our data expand $Ra$ by roughly a decade and $E$ by a factor of three compared to previous laboratory studies in water (and with significantly reduced influence from centrifugal acceleration), allowing us to make closer comparisons to asymptotic studies -- and geophysical settings -- than before. We present measurements of heat transfer and temperature gradients at the mid-height of the fluid layer. Our results demonstrate that asymptotically-predicted transitions between different flow regimes are quantitatively reproduced in laboratory settings. We confirm that the range of ``geostrophic turbulence'' (described below) expands as rotational influence increases. Our data also extend into regions of parameter space where the thermal forcing relative to rotation is stronger than allowed by the asymptotic equations, i.e. strong enough to weaken the rotational constraint. Here, we observe a novel scaling in the heat transfer and in the mid-height temperature gradient, which could identify the existence of a new scaling regime between the previously studied regimes. Furthermore, we observe that this scaling range becomes more and more prominent as rotation is increased ($E$ is reduced).

\section{Governing parameters}
Regimes of convection are often characterized by relationships between the heat transfer parameters. The Rayleigh number $Ra = \alpha_T g \Delta T H^3 / \nu \kappa$ describes the magnitude of thermal forcing, where $\alpha_T$ is the thermal expansion coefficient, $g$ the gravitational acceleration, $\Delta T$ the temperature difference between upper and lower boundaries, $H$ the height of the fluid layer, $\nu$ the kinematic viscosity, and $\kappa$ the thermal diffusivity. The Nusselt number $Nu=q H/k \Delta T$ describes the heat transfer efficiency, where $q$ is the measured heat flux and $k$ the thermal conductivity of the fluid. $Nu=1$ implies pure thermal conduction while $Nu > 1$ implies convective heat transport. The Prandtl number $Pr = \nu / \kappa$ describes the material properties of the fluid through the ratio of viscous and thermal diffusivities. $Pr$ is often fixed at or near 1 in extreme simulations \cite{Ecke14}, but can range from $\sim 0.1$ to $10$ in planetary settings \cite{Schubert11}.

Rotational influence is represented by the Ekman number $E=\nu / 2 \Omega H^2$, where $\Omega$ is the angular frequency; lower $E$ corresponds to greater rotational constraint. The convective Rossby number $Ro = \left(Ra E^2/Pr \right)^{1/2}$ compares rotational and convective forces: traditionally, a constant value of $Ro \sim \mathcal{O}\left(10^{-1}-1\right)$ marks the transition between ``rotationally-affected'' and``rotationally-unaffected'' convection \cite{Gilman77,Stevens13}. Rotation contributes a vertical `stiffness' to the flow, suppressing the onset of convection to $Ra_C = 8.7 E^{-4/3}$ \cite{Chandra61} and to the form of cellular structures with horizontal scale $\ell_{\nu} \sim E^{1/3}H$ \cite{Zhang00, Stellmach04}. In both nonrotating and rotating convection, the thermal forcing relates to the heat transfer efficiency via power law scalings: $Nu \sim Ra^\gamma$ (or, similarly, $Nu \sim \left(Ra/Ra_C\right)^{\gamma'}$), with distinct values of $\gamma$ (or $\gamma'$) for distinct behavioral regimes \cite{Malkus54,Ahlers09}.

\section{Experimental methods}

\subsection{Experimental apparatus}
\begin{figure}
\centering
\includegraphics[width=\textwidth]{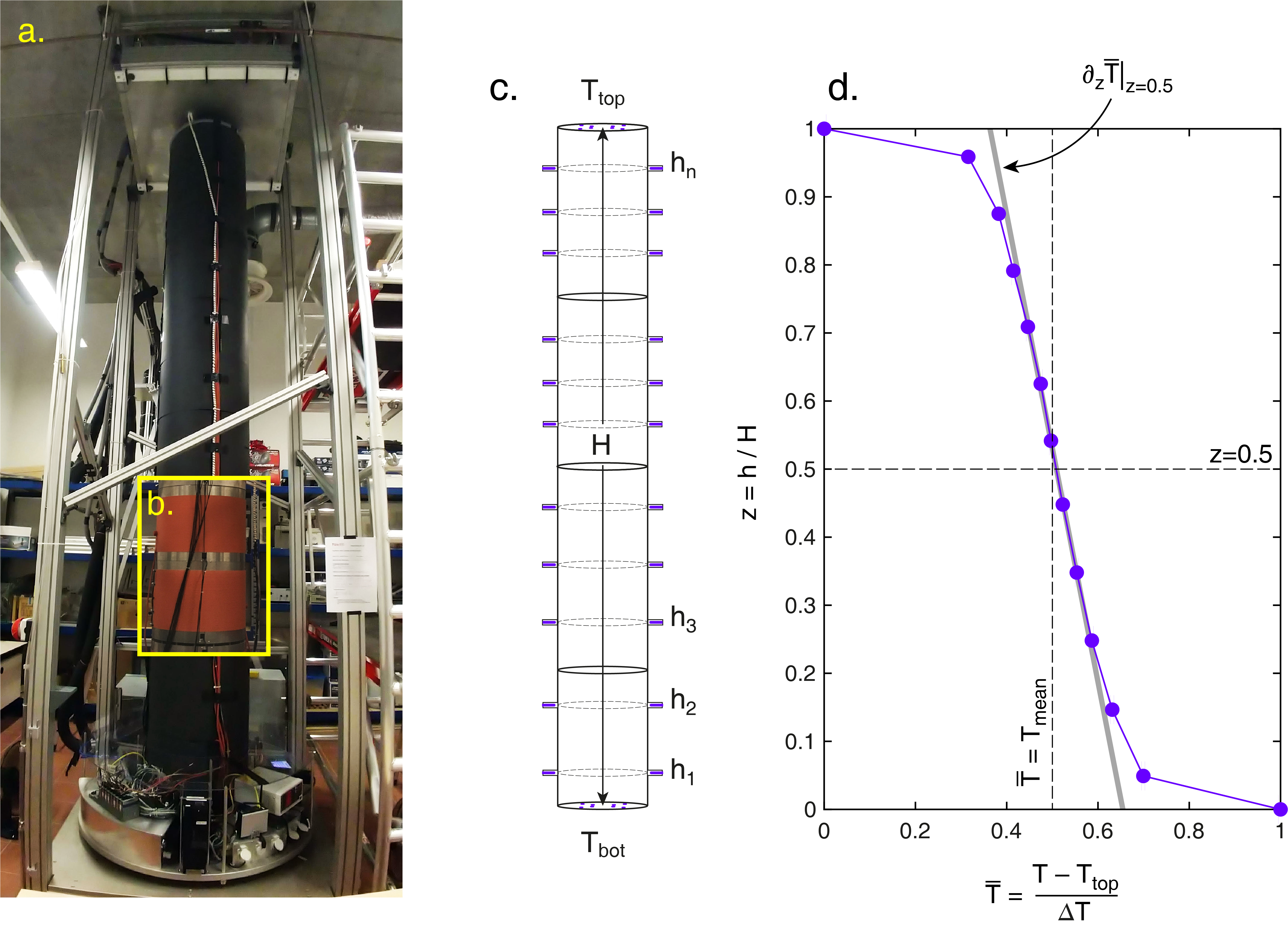}
\caption{\label{F:setup}(a) A photograph of the 4 m high rotating convection setup TROCONVEX. A several centimeter thick layer of black insulating foam covers the experimental vessel. (b) A cutout showing the strip heaters underneath the outermost layer of foam. These heaters match the temperature of the fluid layer to minimize sidewall thermal losses. (c) Sidewall thermistor placements at heights $h_1$, $h_2$, ... $h_n$. (d) Example of a normalized temperature profile from the $E=1.00 \times 10^{-8}$, $Ra=8.66 \times 10^{12}$ case. The temperatures at each height are averaged over time and space. The quantity $\midgrad$ is derived from a linear fit through the middle six temperatures.}
\end{figure}

TROCONVEX (Fig.~\ref{F:setup}) is a rotating convection apparatus designed and constructed at Eindhoven University of Technology. The experimental vessel is a Lexan cylinder of height $H$ up to $4$ m and constant diameter $D=0.39$ m, filled with water and held between copper plates which form the upper and lower boundaries. The mean temperature of the water is kept close to $31^\circ\mathrm{C}$, corresponding to $Pr \approx 5.2$ (see Table~\ref{T:expt} in Appendix \ref{ch:app_table} for exact conditions per run). The adverse temperature gradient is imposed at the top and bottom boundaries: Heat is passed into the system through the bottom plate by an electrical resistance heater and extracted from the top plate via a water-cooled heat exchanger. The heater, connected to a custom-made power supply, supplies between 4 and 1200 W to the system for this study. The top heat exchanger consists of a double-wound spiral passage through which water recirculates from an external coupled Thermoflex 2500 chiller/Sahara AC200 thermostated bath system. The Lexan cylinder is split up vertically into multiple segments, allowing us to use $H=0.8$, $2.0$ and $4.0$ m tank heights (aspect ratio $\Gamma = D/H \approx 1/2$, $1/5$ and $1/10$, respectively) to cover broader parameter ranges. We conduct many experiments that overlap in parameter values at different tank heights (see Table~\ref{T:expt} in Appendix \ref{ch:app_table}) which have, thus far, shown no significant effect of aspect ratio $\Gamma$ on the heat transfer.

As is apparent in Fig.~\ref{F:setup}(d), the bulk region of the fluid can contain significant temperature gradients. In order to control heat losses through the sidewall with precision, each segment is further subdivided vertically into two or three sections [Fig.~\ref{F:setup}(b)], each of which is wrapped in insulating foam and covered with an aluminium heat shield in contact with a flexible resistance strip heater (eleven sections total at $\Gamma = 1/10$). Two negative temperature coefficient (NTC) resistance thermistors are placed at the mid-height of each section, opposite one another (i.e., separated by $180^\circ$), to measure the temperature of the fluid layer; their average temperature is then used as a set point for the corresponding heater. Layers of insulation separate and surround all of these components such that the heaters only passively follow the set point without actively heating the fluid layer. Losses through the Lexan sidewall are therefore minimized even in the presence of large vertical temperature gradients. The sidewall thermistors also turn out to be an important diagnostic tool, described in Sec.~\ref{ch:meas_tech}.

The experimental vessel, power supply, and measurement instruments and controllers are all mounted on the rotating table. Rotation is powered by a Lenze geared motor and gearbox which mesh with a large gear fixed under the table. This allows for precise control of the rotation rate, necessary for maintaining specific Ekman numbers $E$. For the purpose of this study, centrifugal effects are an unwanted externality, and we seek to reduce them as much as possible. Centrifugation is characterized by the Froude number, $Fr = \Omega^2 D/2g$, the ratio between centrifugal forcing at sidewall of the tank and gravitational acceleration. We maintain $Fr < 0.12$ throughout all cases (see Table~\ref{T:expt} in Appendix \ref{ch:app_table}), meaning gravity is roughly an order of magnitude stronger than centrifugal acceleration even at the sidewall. \citet{Horn19} argue, though, that centrifugation is significant as long as $Fr \gtrsim \Gamma/2$, with simulation results showing the temperature gradient at the sidewall experiencing much greater vertical asymmetry than that measured in the bulk. The $E=10^{-8}$ cases at $\Gamma = 1/10$ then exceed this limitation by a factor of two. However, temperature profiles (such as the example shown in Fig.~\ref{F:setup}d, where $Fr = 0.115$) lack the strong vertical asymmetry predicted by \cite{Horn19}, and we believe that centrifugation plays a minimal role in our cases.

\subsection{Measurement techniques\label{ch:meas_tech}}
The results in this work are mainly derived from temperature measurements by NTC resistance thermistors installed in the top, bottom, and sidewall boundaries of the vessel, within 0.7 mm away from the fluid layer [Fig.~\ref{F:setup}(c)]. The top and bottom boundaries have identical configurations of eight thermistors spaced out horizontally across the copper plates, while the sidewall thermistors are arranged two per height at heights $h_1$ through $h_n$. The number of sidewall measurements therefore depends on $\Gamma$, with $n=2$, 5, and 11 measurement heights for the $\Gamma = 1/2$, 1/5, and 1/10 tanks, respectively. Measurements are taken every second, simultaneously across all thermistors, for $N =\mathcal{O}\left(10^{4}-10^{5}\right)$ timesteps. Global heat transfer parameters $Nu$ and $Ra$ rely on our measurements of $\Delta T = T_\text{bot} - T_\text{top}$, where $T_\text{bot}$ and $T_\text{top}$ are the horizontally averaged temperatures at the bottom and top boundaries, respectively. We calculate the fluid properties of water based on the average temperature, $T_\text{mean} = \left(T_\text{bot}+T_\text{top}\right) / 2$, and using formulae from \citet{Lide03}.

The heat flux through the fluid layer, needed to calculate the Nusselt number $Nu$, is measured by a Hioki PW3335 power meter. A small amount of heating power is lost from the other side of the heat pad and the sidewalls and dissipated to the room instead of contributing to heating the fluid layer. We estimate these heat losses by manually inputting a temperature setpoint of $31^\circ\mathrm{C}$ at top, bottom, and sidewall of the fluid layer, and observing the amount of heating power that must be supplied by the bottom heater to maintain this state. This quantity (about 0.6 W for the 2 m tank setup, and estimated to be twice that for the 4 m setup) is then subtracted from the total heating power for calculating $Nu$. The analysis of the statistical error in quantities measured over long times is detailed in Appendix \ref{ch:app}.

As with the top and bottom boundary temperatures, sidewall temperatures are averaged over time and between the (two) thermistors at the same height $h$. The raw temperature profile for each case can be written $\left[T_\text{bot},~T(h_1),~T(h_2),~...,~T(h_n),~T_\text{top}\right]$. We normalize this profile as
\begin{equation}
\label{eq:grad}
\overline{T} = \frac{T-T_\text{top}}{\Delta T} \, .
\end{equation}
The mid-height temperature gradient $\midgrad$ is then defined as a best-fit linear trend of $\overline{T}$ versus $z$ for temperatures in the vicinity of $z=0.5H$: we use $\overline{T}(h_1,~h_2)$ for $\Gamma=1/2$, $\overline{T}(h_1,~h_2,~...,~h_5)$ for $\Gamma = 1/5$, and $\overline{T}(h_3,~h_4,~...,~h_8)$ for $\Gamma = 1/10$.

\begin{figure}
\centering
\includegraphics[width=0.48\linewidth]{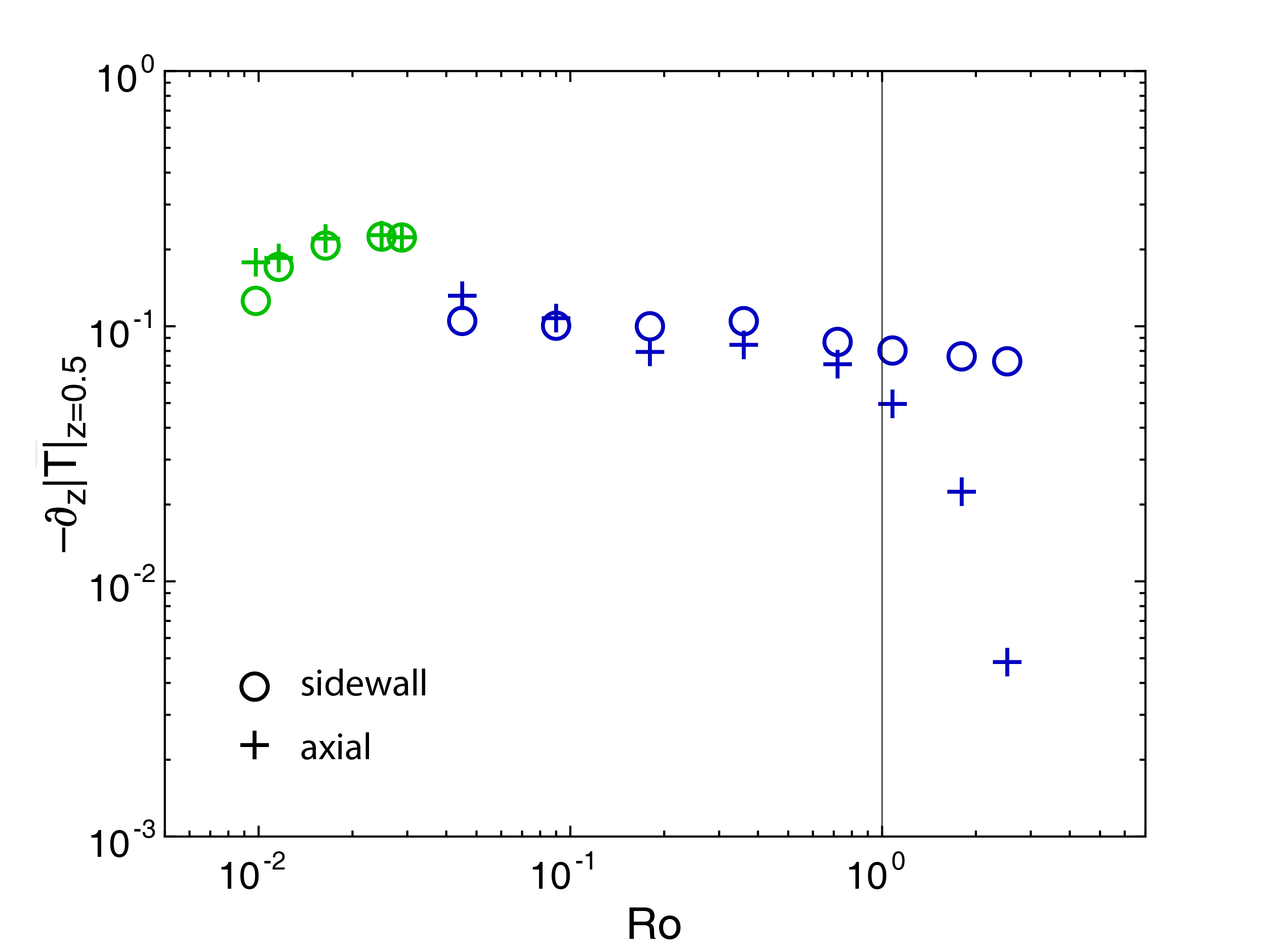}
\caption{\label{F:wall}The mid-height temperature gradient (\ref{eq:grad}) at the wall ($r=D/2$, circles) and at the central axis ($r=0$, crosses) in cylindrical rotating convection DNS with non-slip boundary conditions. Data from \cite{Kunnen10} (blue symbols): Ekman numbers $3.6 \times 10^{-6} \leq E \leq 2.0 \times 10^{-4}$ at constant $Ra=10^{9}$ and $Pr=6.4$. Data from \cite{deWit20} (green symbols): Rayleigh numbers $5.0\times 10^{10} \leq Ra \leq 4.3\times 10^{11}$ at constant $E=10^{-7}$ and $Pr=5.2$. The sidewall and axial temperature gradients match closely for large part of the $Ro$ range. They start to diverge only when rotation begins to dominate (for $Ro\lesssim 10^{-2}$ in \cite{deWit20}) or when rotational influence is mostly lost (for $Ro \gtrsim 1$ in \cite{Kunnen10}).}
\end{figure}

We note that previous studies in nonrotating convection report different temperature profiles at the sidewall versus the axial region of the vessel \cite{Brown07b}. This is not the case, however, in rotating convection: in Fig.~\ref{F:wall} we plot $\midgrad$ versus $Ro$ for cylindrical rotating DNS with non-slip boundaries conducted in \cite{Kunnen10} and \cite{deWit20}. Unlike nonrotating convection, the sidewall and axial temperature gradients match closely when rotation affects the flow ($Ro_C \lesssim 1$) \cite{Gilman77}. Only when rotation is very dominant over convection, in the columnar flow regime, do we see axial and sidewall gradients diverge a bit again (at $Ro\lesssim 10^{-2}$ for \cite{deWit20}). Apart from a few data points, all our rotating data exist in the intermediate range where the gradients are equal irrespective of the radial position.

Recent simulations and experiments have focused on a newly discovered property of the sidewall region: the persistent presence (beyond steady onset) of so-called wall modes in confined rotating convection \cite{deWit20,Favier20,Shishkina20,Zhang20}. These must be discussed in the analysis of our data. The wall modes complicate quantitative comparison of heat flux between experiments and DNS, explained in Sec.~\ref{ch:heatxfer}. However, perhaps counterintuitively, they have no effect on our sidewall temperature gradient measurements. This is explained further in Sec.~\ref{ch:tgradmid}.

\section{Results}

\subsection{Flow visualizations}
\begin{figure}
\centering
\includegraphics[width=\linewidth]{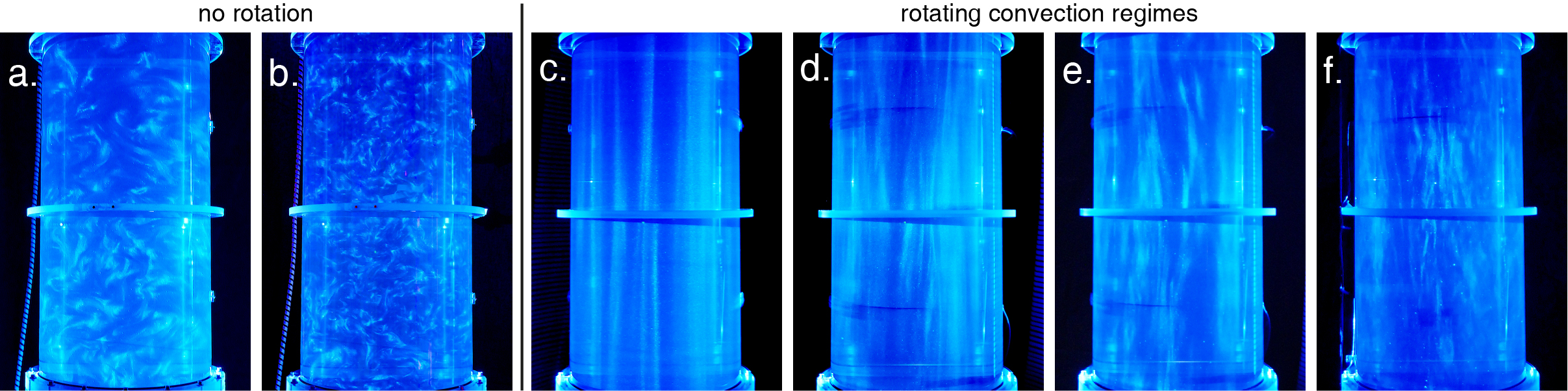}
\caption{\label{F:imgs} \footnotesize Flow field visualizations in a vertical slice of the $\Gamma=D/H=1/5$ tank. Panels (a,b): nonrotating convection at $Ra =$ (a) $1.4 \times 10^{11}$, (b) $2.2 \times 10^{12}$. Panels (c--f): rotating convection at $E = 5 \times 10^{-8}$ and $Ra =$ (c) $9.6 \times 10^{10}$ (convective Taylor columns), (d) $8.6 \times 10^{11}$ (plumes), (e) $1.2 \times 10^{12}$ (geostrophic turbulence), (f) $3.3 \times 10^{12}$ (rotationally-influenced turbulence).}
\end{figure}

We set the stage by presenting visualizations of different flow regimes from the $\Gamma=1/5$ setup (Fig.~\ref{F:imgs}). The water is seeded with neutrally buoyant rheoscopic particles \cite{Borrero18} and illuminated with a vertical light sheet -- regions of strong shear appear as bright or dark streaks. Fig.~\ref{F:imgs}(a,b) show the turbulent flows at high $Ra$ and no rotation. Increasing $Ra$ decreases the scale of flow structures. Under rotation, Fig.~\ref{F:imgs}(c--f), the dominant flow structures evolve differently as $Ra$ increases. In the context of asymptotic simulations where $E, Ro \rightarrow 0$ \cite{JulienGAFD12,Aurnou15}, panel (c) corresponds to ``convective Taylor columns'' ($Ra=2Ra_C$) where narrow structures span the tank vertically, panel (d) to ``convective plumes'' ($Ra=18Ra_C$) where the columns interact laterally and become wavy, and panel (e) to ``geostrophic turbulence (GT)'' ($Ra=30Ra_C$), where convective forces have destroyed the columnar structure but flows remain constrained to rotational length scales. Asymptotic simulations predict that the transition between columns and plumes occurs at \cite{Nieves14}:
\begin{equation}
\label{eq:nieves}
Ra = 55 E^{-4/3} =6.3 Ra_C \, .
\end{equation}
In contrast, no such prediction can be confidently made for the plumes to GT transition: for low $E$ ($\lesssim 10^{-6}$), GT exists at the limit of accessibility for asymptotic simulations and DNS. It remains largely uncharacterized at $Pr \neq 1$ and with poorly constrained scaling properties even at $Pr = 1$ \cite{JulienGAFD12,Ecke14,Stellmach14,Kunnen16}.

The onset horizontal scale $\ell_\nu$ -- believed to accurately describe flows in Fig.~\ref{F:imgs}(c--e) -- serves as a necessary condition for deriving the asymptotically-reduced equations \cite{Sprague06}. This assumption persists until buoyancy takes over the horizontal length scale, theorized to occur at \cite{JulienPRL12,Gastine16}:
\begin{equation}
\label{eq:asymp}
Ra \sim E^{-8/5}Pr^{3/5} \, .
\end{equation}
Fig.~\ref{F:imgs}(f) ($Ra=70Ra_C$) lies beyond this upper bound and demonstrates a remarkably different flow morphology. At first glance, flows here in the range of $Ra > E^{-8/5}Pr^{3/5}$ and $Ro \lesssim 1$ are well-studied \cite{Zhong09,Stevens13}. Importantly, though, these studies all occur at moderate $E$ values, and the flow properties undergo fundamental changes as $E$ decreases: for example, $Nu$ overshoots above $Nu_0$ in this range for $E\gtrsim 10^{-6}$ \cite{Zhong09,Stevens13}, but becomes suppressed below $Nu_0$ at $E \lesssim 10^{-7}$ \cite{Cheng15}. The behavior seen in Fig.~\ref{F:imgs}(f), then, begs further examination. Our experiment is well-suited for this purpose due to its ability to explore high $Ra/Ra_C$ values at simultaneously low $E$ \cite{Cheng18} -- specifically, higher $Ra/Ra_C$ than is possible in simulations of asymptotically reduced equations. 

\subsection{\label{ch:heatxfer}Heat transfer}
\begin{figure}
\centering
\includegraphics[width=\textwidth]{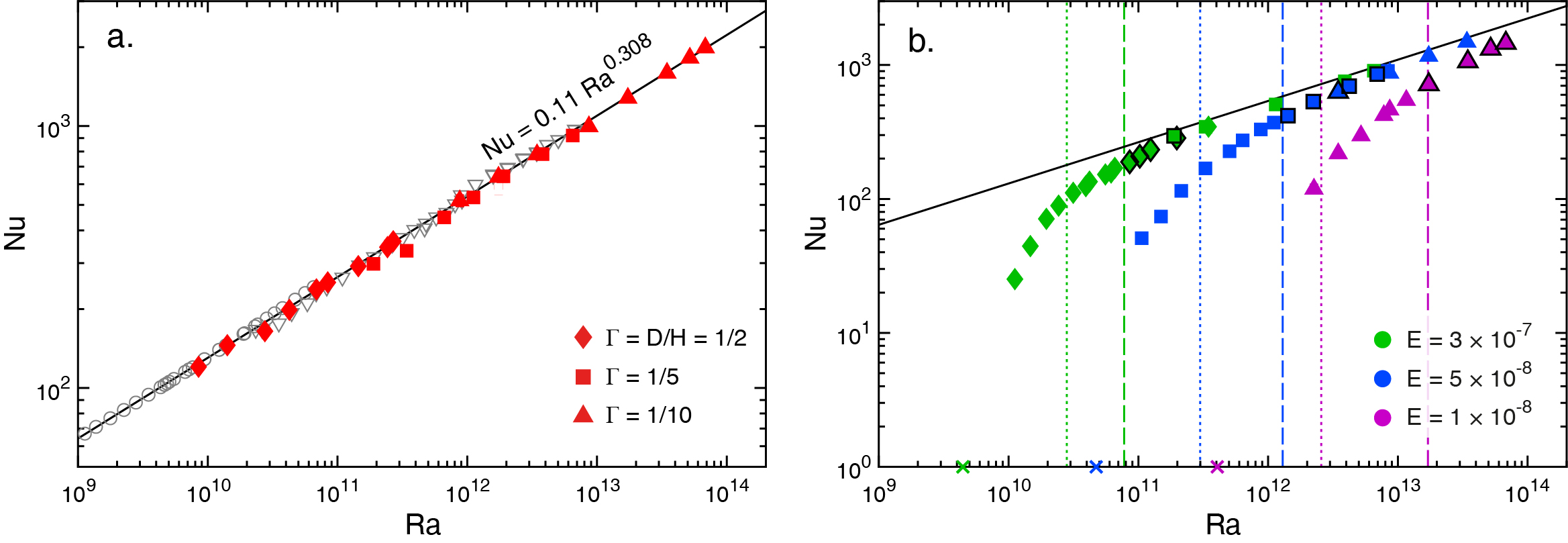}
\caption{\label{F:NuRa} \footnotesize Nusselt number ($Nu$) plotted versus Rayleigh number ($Ra$) for (a) nonrotating and (b) rotating convection experiments. In both panels, marker shape represents tank 
aspect ratio $\Gamma$; the solid line our nonrotating convection fit (\ref{eq:RBC}). Data from previous studies are included as open grey circles \cite{Funf05} and triangles \cite{Cheng15}. In panel (b), color represents Ekman number ($E$), $Ra_C$ in each case is marked by `$\times$,' and black-outlined symbols are within our new hypothesized scaling range (to be precisely defined later). Dotted lines represent the columns--plumes transition (\ref{eq:nieves}) from \cite{Nieves14}; dashed lines represent the transition (\ref{eq:asymp}) where the asymptotically reduced equations are projected to break down, it is the flow transition from GT to the hypothesized new scaling range.}
\end{figure}

To quantify the visualized regimes, we first examine scalings between the heat transfer parameters. Before considering rotating convection, we validate our results with nonrotating turbulent heat transfer. Fig.~\ref{F:NuRa}(a) shows our nonrotating data in terms of $Nu$ versus $Ra$. We achieve a maximum $Ra=7 \times 10^{13}$ -- nearly a decade higher than any previous $Pr > 1$ study. The data follow a best-fit scaling
\begin{equation}
\label{eq:RBC}
Nu_0 = 0.11 \left(^{+0.02}_{-0.01}\right) Ra^{0.308 \pm 0.005} \, ,
\end{equation}
agreeing with previous scaling exponents $\sim 0.3-0.33$ found in water \cite{Brown05b,SunJFM05,Cheng15} and other fluids \cite{Niemela00,Funf09,Niemela10}. They are also consistent with the classical prediction $Nu \sim Ra^{1/3}$ where the bulk is sufficiently turbulent as to be approximately isothermal, with the temperature gradient confined to the upper and lower thermal boundary layers \cite{Malkus54}. There is no evidence of transition to a steeper scaling that would indicate `ultimate' convection: the state of convection where the boundary layers have become turbulent \cite{Kraichnan62,Ahlers09} with more efficient heat transfer characterized by a steeper scaling exponent $\gamma>1/3$ in the scaling relation $Nu\sim Ra^\gamma$. The transition location $Ra\sim 10^{11}$--$10^{13}$ where ultimate convection sets in is disputed \cite{Chavanne97,He12}. Note that our comparatively narrow geometry ($\Gamma = 1/10$) is possibly detrimental to the transition to ultimate convection; recent numerical simulations in a similar geometry \cite{Iyer20} also do not observe a transition.

We present our rotating convection heat transfer data as $Nu(Ra)$ graphs at three different Ekman numbers, $E=10^{-8}$, $5\times 10^{-8}$ and $3\times 10^{-7}$. Each curve [Fig.~\ref{F:NuRa}(b)] follows a characteristic succession of ever shallower slopes as $Ra$ increases (as postulated in \cite{Cheng18}), eventually merging with the nonrotating scaling (\ref{eq:RBC}). As is characteristic for low $E$, rotational $Nu$ values lie below the nonrotating $Nu_0$ values until well beyond onset. This separation becomes more pronounced as $E$ decreases: the $E=3 \times 10^{-7}$ trend first comes within 10\% of the nonrotating trend at $Ra \approx 40 Ra_C$, while for $E = 5 \times 10^{-8}$ this does not occur until $Ra \approx 400Ra_C$. This confirms that the so-called geostrophic regime of rotation-dominated convection, with its reported subdomains as cells, columns, plumes and GT, expands as $E$ is lowered.

It is technically difficult to connect our lowermost experimental data point at $Ra=1.12 \times 10^{10}$, $E=3\times 10^{-7}$ to the onset of convection (green `$\times$' on the horizontal axis); the minute temperature differences required cannot be stably maintained \cite{King12,Cheng18}. But it is clear that a steep $Nu(Ra)$ scaling in the cellular and columnar regimes is required. This picture is consistent with direct numerical simulation (DNS) studies at low $E$ and $Pr \simeq 3$--$10$ \cite{King12,Stellmach14,Cheng15} that do observe steep $Nu(Ra)$ scaling. However, there is an important difference in domain composition: these studies typically use rectilinear domains with periodic boundary conditions in the horizontal directions rather than a cylinder. Sidewall modes \cite{deWit20,Favier20,Shishkina20,Zhang20} are then present in the laboratory experiments but not in DNS. The wall modes lead to increased $Nu$, with the effect more pronounced at lower $Ra/Ra_C$ \cite{deWit20}. We therefore refrain from making quantitative comparisons of overall heat transfer between experiments and DNS.

\begin{figure}
\centering
\includegraphics[width=\linewidth]{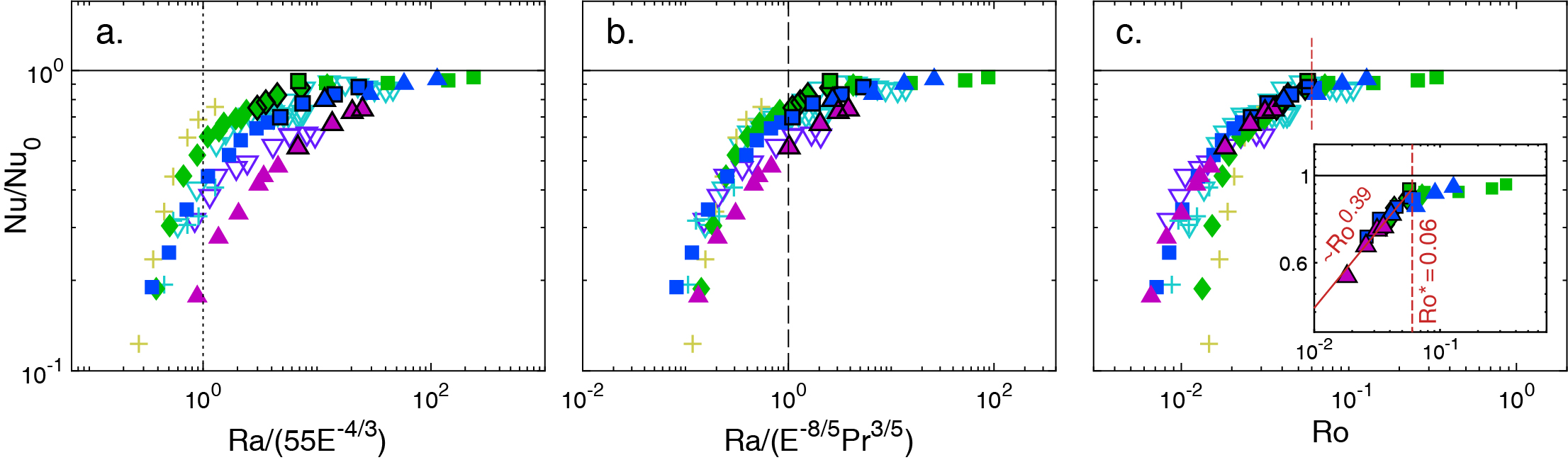}
\caption{\label{F:comp} $Nu$ compensated by nonrotating scaling $Nu_0$ as per Eq.~(\ref{eq:RBC}) versus: (a) $Ra/55 E^{-4/3}$ as per Eq.~(\ref{eq:nieves}) \cite{Nieves14}; (b) $Ra/(E^{-8/5}Pr^{3/5)}=Ra/(2.7E^{-8/5})$ as per Eq.~(\ref{eq:asymp}) \cite{JulienPRL12}, and (c) $Ro$ \cite{Gilman77}. Bullet color and shape are the same as in Fig.~\ref{F:NuRa}, with additional points from \cite{Cheng15}: numerical $E=1 \times 10^{-6}$ cases are yellow crosses, numerical $E=1 \times 10^{-7}$ cases are cyan crosses, lab $E \simeq 10^{-7}$ cases are empty cyan triangles, and lab $E \simeq 3 \times 10^{-8}$ cases are empty indigo triangles. The inset in panel (c) shows our rotating data above Eq.~(\ref{eq:asymp}). A change in slope occurs at $Ro^* \approx 0.06$, which we argue is the upper limit of the new scaling range.}
\end{figure}

We anticipate that the different regimes of flow phenomenology can be recognized as specific power-law scaling ranges with characteristic scaling exponents~\cite{Cheng18}. In Fig.~\ref{F:comp}, though, we illustrate that the $Nu(Ra)$ scaling alone does not provide conclusive evidence for regime transitions. Here $Nu$ is compensated by the nonrotating $Nu_0$ fit (\ref{eq:RBC}), and $Ra$ is compensated by several transition arguments. In contrast to previous studies \cite{Ecke14,Cheng15}, none of these arguments definitively collapse data across multiple $E$ values in our extreme parameter range. Compensating $Ra$ by Eq.~(\ref{eq:nieves}) gives little evidence of collapse [Fig.~\ref{F:comp}(a)]. Compensating with Eq.~(\ref{eq:asymp}) collapses the near-onset trends while inducing spread in the GT range [Fig.~\ref{F:comp}(b)]; plotting as a function of $Ro$ does the opposite [Fig.~\ref{F:comp}(c)]. In the latter panel there is satisfactory collapse of our higher $Ro$ data, with a clear transition at~$Ro^*=0.06$ (see inset) that is independent of aspect ratio $\Gamma$. 

We posit that $Ro^*=0.06$ could mark the upper boundary of a new scaling range. The transition point is notably distinct from the well-documented transition from rotation-affected to nonrotating convection \cite{WeissPRL10}: the latter transition has been shown to be strongly dependent on $\Gamma$ and the predicted $Ro_T$ ($Ro_T=\{1.2, 0.40, 0.16\}$ for $\Gamma=\{1/2,1/5,1/10\}$, respectively) is significantly higher than $Ro^*=0.06$. Thus, we anticipate that for $Ro^*<Ro<Ro_T$ we have recovered rotation-affected convection where rotation does not significantly affect the heat transfer scaling; the data points approach the nonrotating $Nu(Ra)$ curve as $Ra$ increases. We note that a qualitatively similar collapse is found in \cite{Ecke14}, but at a significantly higher transition value~$Ro^*=0.35$. The most important difference is the working fluid and associated Prandtl number; Ecke \& Niemela \cite{Ecke14} used low-temperature helium with $Pr=0.7$ while we use water at $Pr=5.2$.

We argue in Sec.~\ref{ch:tgradmid} that the lower boundary of the new scaling range is given by Eq.~(\ref{eq:asymp}): $Ra\ge E^{-8/5}Pr^{3/5}$. An indicative fit of heat transfer in the new regime results in~$Nu/Nu_0\sim Ro^{0.39}$ or~$Nu\sim Ra^{0.52}$ ($Nu\sim Ra^{\gamma^*}$ with $\gamma^*=0.52$).

\begin{figure}
\centering
\includegraphics[width=0.8\linewidth]{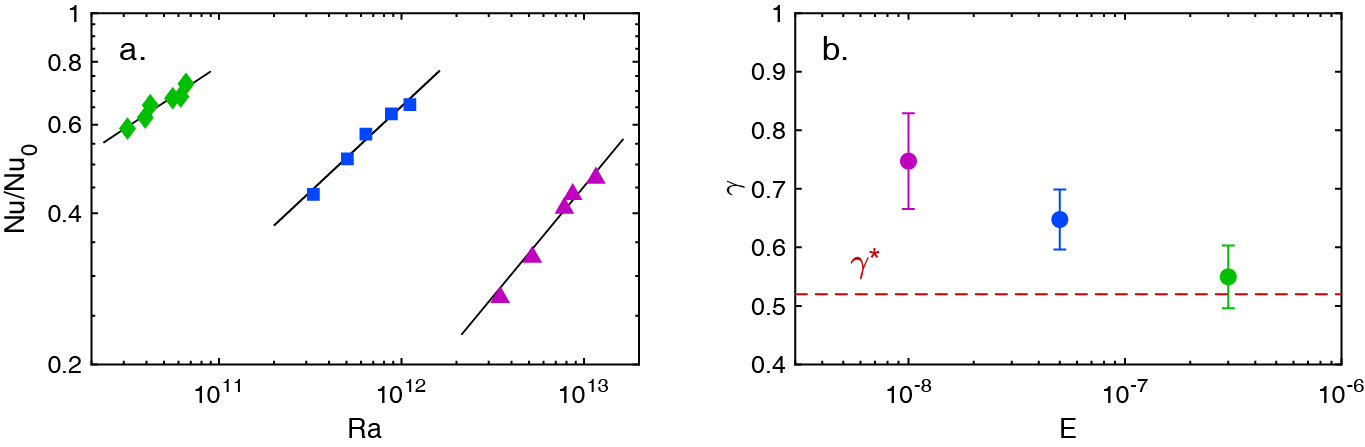}
\caption{\label{F:zoom}(a) Zoomed-in view of the $Nu$ data points of Fig. \ref{F:NuRa}(b) in the plumes/GT range, plotted as $Nu/Nu_0$ as a function of $Ra$ for clarity. Weighted least-squares power-law fits are included with solid black lines. Bullet color and shape as in Fig. \ref{F:NuRa}. (b) Corresponding $Nu(Ra)$ scaling exponents $\gamma$ as a function of $E$. The error bars indicate one standard deviation error intervals resulting from the weighted least-squares fits. The red dashed line indicates $\gamma^*=0.52$, the previously discussed scaling exponent of the new scaling range beyond the GT regime.}
\end{figure}

Additionally, we can consider heat transfer in the regimes characterized by plumes and geostrophic turbulence, the range between dotted and dashed lines in Fig. \ref{F:NuRa}(b). These data points also display an approximate power-law scaling $Nu\sim Ra^\gamma$; the scaling exponent $\gamma$ is generally larger than for the previously discussed new scaling range and shows a clear dependence on $E$. A close-up view of the data points in this range is plotted in Fig. \ref{F:zoom}(a), where we plot $Nu/Nu_0$ as a function of $Ra$ to have the points lie closer together. Solid black lines are weighted least-squares power-law fits. Fig. \ref{F:zoom}(b) presents the corresponding scaling exponents $\gamma$ as a function of $E$. Though the range is not extensive, it is clear that steeper scaling is observed for smaller $E$. This is again in line with the projected succession of scaling ranges with descriptive exponents per flow state \cite{Cheng18}. Compared to the reported asymptotic scaling $\gamma =1.5$ for geostrophic turbulence \cite{JulienPRL12} our exponents are small, but increasing as $E$ is reduced.

\subsection{\label{ch:tgradmid}Mid-height temperature gradient}
In lieu of further transition information from globally-averaged parameters, we shift focus to the time-averaged vertical temperature gradient $\midgrad$. In nonrotating turbulent convection, the temperature profile is sharply divided between the bulk, which is nearly isothermal ($-\partial_z \overline{T} \approx 0$), and the bottom and top thermal boundary layers, within which nearly all of the temperature drop $\Delta T$ occurs \cite{Grossmann00,Ahlers09}. In rotating convection, however, the shape of the temperature profile evolves as $Ra/Ra_C$ changes and with it the flow morphology, giving it diagnostic properties \cite{JulienGAFD12,Nieves14,Kunnen16}. The aforementioned wall modes of confined convection \cite{deWit20,Favier20,Shishkina20,Zhang20} actually do not affect long-time averaged temperature values; their symmetrical structure (azimuthal distribution into a `colder' and a `warmer' half) and gradual azimuthal precession \cite{deWit20} make it so that the wall mode signature is averaged out in the sidewall thermistor time series. Hence the time-averaged sidewall gradient still nicely follows that of the bulk, independent of the presence of the wall mode (in the manner of the DNS comparison in Fig.~\ref{F:wall}).

\begin{figure}
\centering
\includegraphics[width=1\linewidth]{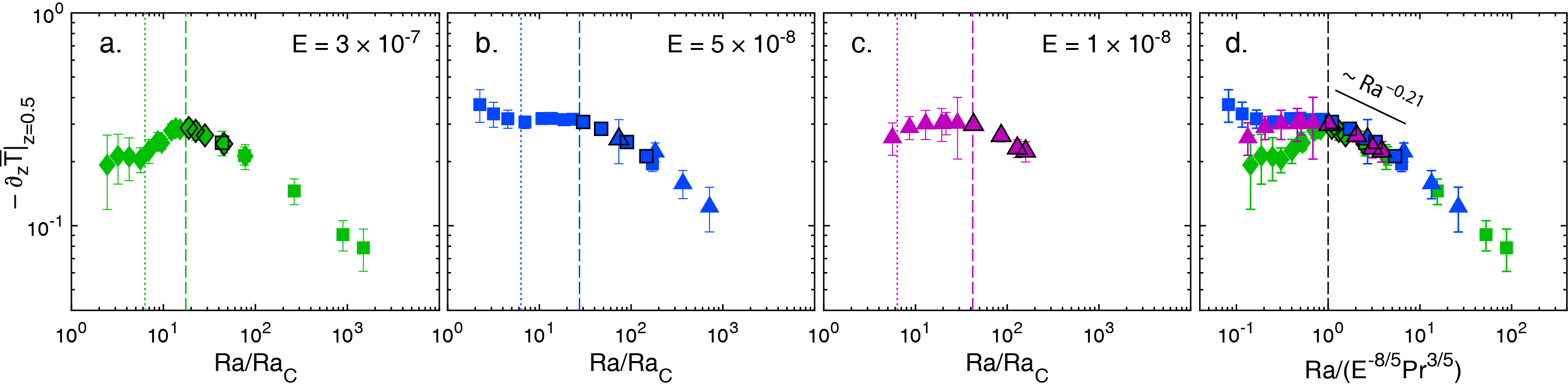}
\caption{\label{F:grad} Normalized mid-height temperature gradient $\midgrad$ versus $Ra/Ra_C$ for (a) $E=3 \times 10^{-7}$, (b) $E=5 \times 10^{-8}$, and (c) $E=1 \times 10^{-8}$. Symbol shapes represent $\Gamma$ as in Fig.~\ref{F:NuRa}. Dotted lines represent Eq.~(\ref{eq:nieves}); dashed lines Eq.~(\ref{eq:asymp}). (d) Normalized gradient $\midgrad$ versus $Ra/(E^{-8/5}Pr^{3/5})=Ra/(2.7E^{-8/5})$ as per Eq.~(\ref{eq:asymp}), for all three $E$ values. A fit to the black-bordered points (those lying in the new scaling range) across all $E$ values gives $\midgrad \sim Ra^{-0.21}$.}
\end{figure}

Fig.~\ref{F:grad}(a--c) show that the temperature gradient measured at mid-height, $\midgrad$, is indeed a robust tool for determining regime transitions at each $E$ value. Remarkably, the transitions found in asymptotic simulations show excellent quantitative agreement with the transitions we observe in our measured $\midgrad$, regardless of differences in domain composition and boundary conditions. In the cellular and columnar regimes, increasing $Ra$ leads to a decreasing temperature gradient. At the transition point described by Eq.~(\ref{eq:nieves}) this trend reverses, indicating that increasing $Ra/Ra_C$ now forces more of the temperature gradient into the interior as the horizontal rigidity of bulk flow structures relaxes. Visualizations of the flow and $Nu(Ra)$ data do not reveal where the plumes--GT regime transition takes place, but asymptotic studies posit that GT corresponds to $\midgrad$ flattening with increasing $Ra/Ra_C$ \cite{JulienGAFD12}. Our results do appear to manifest such a flattening at $Ra/Ra_C \simeq 15$ at each $E$ value. This abruptly gives way to a decreasing trend at the transition point described by Eq.~(\ref{eq:asymp}), which could be considered the lower boundary of the new scaling regime that we postulate. A factor $E^{-4/15}$ separates Eqs.~(\ref{eq:asymp}) and (\ref{eq:nieves}), identifying that the parameter range of plumes/GT expands as $E$ decreases toward geophysical values.

The trends identified in Fig.~\ref{F:grad}(a--c) rely on careful consideration of the error bars in $\midgrad$, which we detail in Appendix \ref{ch:app}. The error can be seen to increase towards smaller values of $Ra/Ra_C$. This is mainly due to the smaller $\Delta T$ values used there with correspondingly larger relative error; $\Delta T$ is used for normalization in $\midgrad$. Furthermore, the error also grows towards small values of $\midgrad$ that are found at the largest $Ra/Ra_C$ values. There, the temperature difference over the bulk reduces, with larger relative error as a result.

In Fig.~\ref{F:grad}(d) we plot $\midgrad$ data for all $E$ values versus $Ra$ rescaled by Eq.~(\ref{eq:asymp}) using $1$ as prefactor. Overplotting separate $E$ trends collapses all data in this regime into an approximate $\midgrad \sim Ra^{-0.2}$ scaling, in sharp contrast to the $Ra^0$ scaling -- and lack of collapse over multiple $E$ values -- reported for the traditional rotationally-affected regime (see Table 2 of Ref. \cite{King13}, the authors labeled  it the `weakly rotating' regime). We thus postulate Eq.~(\ref{eq:asymp}) as the lower bound of a novel scaling range that displays temperature statistics different from previously identified scaling ranges: both ``rotationally-affected convection" and GT. Data points belonging to the new scaling range are marked by black outlines in Figs.~\ref{F:NuRa}--\ref{F:grad}. The scaling exponent of $\midgrad$ in this range is a novel result that we cannot currently explain. We speculate that nonrotating-style thermal boundary layers have formed beyond this transition, such that increasing $Ra$ causes stronger mixing in the bulk and isolates more of the temperature gradient into the boundary layers. We are preparing flow measurements using stereoscopic particle image velocimetry, to gain more insight into the flow statistics in different flow regimes and, in particular, to formulate a better understanding of the flow in the proposed new scaling range.

\section{Conclusion}
Our rotating convection survey demonstrates the emergence of several distinct regimes as $E$ is pushed lower than any previous study in water. Scalings between $Nu$ and $Ra$ show consistency with previous results at moderate parameter values while extending them to more extreme values. To determine precise transition locations, the mid-height temperature gradient $\midgrad$ serves as a robust tool: transitions nigh invisible in $Nu(Ra)$ plots [Figs.~\ref{F:NuRa}~\&~\ref{F:comp}] are expressed as pronounced changes in the $\midgrad$ vs. $Ra/Ra_C$ trend [Fig.~\ref{F:grad}]. Our data confirm that the GT range expands as $E$ decreases. They also suggest the existence of an additional, previously unidentified scaling regime just beyond GT, for $Ra > E^{-8/5} Pr^{3/5}$, or $Ro > \left(E/Pr\right)^{1/5}$, and $Ro<Ro^*=0.06$. This range, which we refer to as rotationally-influenced turbulence (RIT), displays behaviors differentiating it from both the GT regime at asymptotically small $E$ \cite{JulienGAFD12} and the rotationally-affected regime at moderately large $E$ \cite{Zhong09,Stevens13}: temperature gradients collapse as $\midgrad \sim Ra^{-0.21}$ and heat flux as $Nu/Nu_0\sim Ro^{0.39}$. In the regime of plumes and GT, we observe that the heat flux scaling is generally steeper than that, with increasing exponent as $E$ is reduced. Furthermore, we show that the mid-height temperature gradient can be used to identify regime transitions in remarkable quantitative agreement with transition predictions from asymptotic simulations, despite different domain composition and boundary conditions.

These results show that the parameter space of turbulent convection, from nonrotating to asymptotically rapid rotation, may show even richer subdivisions than known so far. We argue that there are signs of a new intermediate RIT regime separating ``rotationally-dominated'' convection (GT) from traditional ``rotation-affected convection". Extrapolating the scaling arguments supported in this paper, the spaces between transitions widen as $E$ decreases: the plumes/GT range expands as $Ra \sim E^{-4/15}$ while the proposed new scaling range expands as $Ra \sim E^{-2/5}$. In the geophysical context, estimates for planetary fluid layers give $E \sim 10^{-19}-10^{-12}$ \cite{Schubert11} and $Ra/Ra_C \sim 10^2-10^3$ \cite{ChengAurnou16}. Rotating convection in these layers then invariably inhabits either geostrophic turbulence or turbulence exhibiting the new scaling. Large-scale laboratory experiments are uniquely suited to bridging the gap between asymptotic studies and direct numerical simulations, while simultaneously exploring the parameter space currently out of reach for both simulation approaches. They are thereby instrumental for informing future work on geophysical and astrophysical flows.

\begin{acknowledgments}
The authors have received funding from the European Research Council (ERC) under the European Union's Horizon 2020 research and innovation programme (Grant agreement No. 678634). We are grateful for the support of the Netherlands Organisation for Scientific Research (NWO) for the use of supercomputer facilities (Cartesius) under Grants No. 15462, 16467 and 2019.005.
\end{acknowledgments}

\appendix

\section{Table of experimental data\label{ch:app_table}}

\begingroup
\def\arraystretch{1.2}
\begin{longtable*}{@{\extracolsep{10pt}}cccllcclc}
\caption[]{Experimental data from TROCONVEX. $\Gamma = 0.494, 0.195$, and $0.097$ cases are from the 0.8 m, 2 m, and 4 m high setups, respectively. Nonrotating cases are listed first. Some cases do not have $\midgrad$ values because they were conducted prior to implementation of sidewall temperature measurements.}\\
{$\Gamma$} &{$E$} &{$Fr$} &{$Ra$} &{$Nu$} &{$Pr$} &{$T_\text{mean}$ [$^\circ$C$]$} &{$\Delta T$ [$^\circ$C$]$} &{$\midgrad$} \\ [3pt]
\hline
\endfirsthead
{$\Gamma$} &{$E$} &{$Fr$} &{$Ra$} &{$Nu$} &{$Pr$} &{$T_\text{mean}$ [$^\circ$C$]$} &{$\Delta T$ [$^\circ$C$]$} &{$\midgrad$} \\ [3pt]
\hline
\endhead
    0.494 &{$\infty$} & 0     & 8.49 ($\pm$0.20) E+09 & 121 ($\pm$3)   & 5.99  & 25.48 & 0.85  & 0.052 \\
    0.494 &{$\infty$} & 0     & 1.41 ($\pm$0.02) E+10 & 146 ($\pm$2)   & 6.00  & 25.41 & 1.41  & 0.068 \\
    0.494 &{$\infty$} & 0     & 2.75 ($\pm$0.02) E+10 & 165 ($\pm$1)   & 6.65  & 21.50 & 3.51  &{--} \\
    0.494 &{$\infty$} & 0     & 4.25 ($\pm$0.03) E+10 & 199 ($\pm$1)   & 6.65  & 21.52 & 5.41  &{--} \\
    0.494 &{$\infty$} & 0     & 6.87 ($\pm$0.07) E+10 & 238 ($\pm$2)   & 5.99  & 25.45 & 6.86  & 0.038 \\
    0.494 &{$\infty$} & 0     & 8.41 ($\pm$0.06) E+10 & 253 ($\pm$2)   & 6.73  & 21.10 & 11.00 &{--} \\
    0.494 &{$\infty$} & 0     & 1.45 ($\pm$0.01) E+11 & 292 ($\pm$2)   & 6.75  & 20.99 & 19.07 &{--} \\
    0.494 &{$\infty$} & 0     & 2.43 ($\pm$0.04) E+11 & 345 ($\pm$5)   & 5.73  & 27.21 & 21.97 & 0.035 \\
    0.494 &{$\infty$} & 0     & 2.69 ($\pm$0.03) E+11 & 363 ($\pm$3)   & 6.33  & 23.33 & 30.38 &{--} \\
    0.195 &{$\infty$} & 0     & 1.89 ($\pm$0.04) E+11 & 298 ($\pm$7)   & 5.21  & 31.01 & 0.87  & 0.109 \\
    0.195 &{$\infty$} & 0     & 3.41 ($\pm$0.06) E+11 & 334 ($\pm$6)   & 5.21  & 31.06 & 1.57  & 0.079 \\
    0.195 &{$\infty$} & 0     & 6.64 ($\pm$0.07) E+11 & 448 ($\pm$5)   & 5.21  & 31.01 & 3.06  & 0.069 \\
    0.195 &{$\infty$} & 0     & 1.11 ($\pm$0.01) E+12 & 534 ($\pm$6)   & 5.21  & 31.04 & 5.13  & 0.059 \\
    0.195 &{$\infty$} & 0     & 1.90 ($\pm$0.02) E+12 & 643 ($\pm$7)   & 5.18  & 31.30 & 8.64  & 0.051 \\
    0.195 &{$\infty$} & 0     & 3.80 ($\pm$0.03) E+12 & 781 ($\pm$5)   & 5.21  & 31.03 & 17.49 & 0.045 \\
    0.195 &{$\infty$} & 0     & 6.49 ($\pm$0.06) E+12 & 921 ($\pm$8)   & 5.22  & 30.98 & 29.94 & 0.042 \\
    0.097 &{$\infty$} & 0     & 8.75 ($\pm$0.36) E+11 & 519 ($\pm$27)   & 5.22  & 30.98 & 0.51  & 0.150 \\
    0.097 &{$\infty$} & 0     & 1.74 ($\pm$0.04) E+12 & 639 ($\pm$17)   & 5.22  & 30.98 & 1.01  & 0.118 \\
    0.097 &{$\infty$} & 0     & 3.45 ($\pm$0.97) E+12 & 778 ($\pm$23)   & 5.22  & 30.99 & 1.99  & 0.093 \\
    0.097 &{$\infty$} & 0     & 8.66 ($\pm$0.07) E+12 & 996 ($\pm$19)   & 5.22  & 30.98 & 5.01  & 0.080 \\
    0.097 &{$\infty$} & 0     & 1.73 ($\pm$0.01) E+13 & 1279 ($\pm$9)  & 5.22  & 31.00 & 9.99  & 0.071 \\
    0.097 &{$\infty$} & 0     & 3.46 ($\pm$0.03) E+13 & 1593 ($\pm$15)  & 5.22  & 30.99 & 20.00 & 0.066 \\
    0.097 &{$\infty$} & 0     & 5.19 ($\pm$0.06) E+13 & 1820 ($\pm$20)  & 5.22  & 31.00 & 29.99 & 0.061 \\
    0.097 &{$\infty$} & 0     & 6.83 ($\pm$0.09) E+13 & 1993 ($\pm$25)  & 5.00  & 32.78 & 36.44 & 0.063 \\
    0.494 & 2.88E-07 & 0.092 & 1.12 ($\pm$0.02) E+10 & 25 ($\pm$1)    & 5.22  & 30.94 & 0.84  & 0.193 \\
    0.494 & 2.88E-07 & 0.092 & 1.47 ($\pm$0.02) E+10 & 44 ($\pm$1)    & 5.21  & 31.07 & 1.10  & 0.213 \\
    0.494 & 2.86E-07 & 0.092 & 1.96 ($\pm$0.02) E+10 & 71 ($\pm$1)    & 5.18  & 31.25 & 1.45  & 0.211 \\
    0.494 & 3.00E-07 & 0.106 & 2.43 ($\pm$0.02) E+10 & 89 ($\pm$1)    & 5.92  & 25.91 & 2.36  & 0.205 \\
    0.494 & 2.89E-07 & 0.092 & 3.15 ($\pm$0.02) E+10 & 111 ($\pm$1)   & 5.24  & 30.79 & 2.39  & 0.225 \\
    0.494 & 2.90E-07 & 0.092 & 3.94 ($\pm$0.02) E+10 & 125 ($\pm$1)   & 5.26  & 30.64 & 3.01  & 0.249 \\
    0.494 & 2.98E-07 & 0.106 & 4.20 ($\pm$0.02) E+10 & 135 ($\pm$1)   & 5.86  & 26.28 & 4.00  & 0.245 \\
    0.494 & 2.89E-07 & 0.092 & 5.57 ($\pm$0.02) E+10 & 152 ($\pm$1)   & 5.23  & 30.86 & 4.21  & 0.280 \\
    0.494 & 2.90E-07 & 0.092 & 6.17 ($\pm$0.03) E+10 & 158 ($\pm$1)   & 5.26  & 30.64 & 4.71  & 0.287 \\
    0.494 & 2.99E-07 & 0.106 & 6.59 ($\pm$0.04) E+10 & 171 ($\pm$1)   & 5.90  & 26.02 & 6.36  & 0.284 \\
    0.494 & 2.90E-07 & 0.092 & 8.58 ($\pm$0.06) E+10 & 189 ($\pm$1)   & 5.24  & 30.77 & 6.51  & 0.288 \\
    0.494 & 2.88E-07 & 0.092 & 1.026 ($\pm$0.006) E+11 & 208 ($\pm$1)   & 5.22  & 30.98 & 7.72  & 0.279 \\
    0.494 & 2.97E-07 & 0.106 & 1.242 ($\pm$0.007) E+11 & 233 ($\pm$1)   & 5.85  & 26.41 & 11.75 & 0.266 \\
    0.494 & 3.00E-07 & 0.106 & 1.98 ($\pm$0.01) E+11 & 285 ($\pm$2)   & 5.91  & 25.95 & 19.19 & 0.242 \\
    0.494 & 2.93E-07 & 0.106 & 3.47 ($\pm$0.03) E+11 & 346 ($\pm$3)   & 5.75  & 27.04 & 31.75 & 0.212 \\
    0.195 & 2.99E-07 & 0.002 & 1.89 ($\pm$0.04) E+11 & 296 ($\pm$6)   & 5.23  & 30.92 & 0.87  & 0.245 \\
    0.195 & 3.00E-07 & 0.002 & 3.30 ($\pm$0.04) E+11 & 346 ($\pm$4)   & 5.25  & 30.75 & 1.54  & 0.214 \\
    0.195 & 3.00E-07 & 0.002 & 1.14 ($\pm$0.01) E+12 & 508 ($\pm$6)   & 5.26  & 30.67 & 5.35  & 0.146 \\
    0.195 & 2.99E-07 & 0.002 & 3.90 ($\pm$0.03) E+12 & 756 ($\pm$6)   & 5.23  & 30.85 & 18.10 & 0.091 \\
    0.195 & 2.98E-07 & 0.002 & 6.53 ($\pm$0.07) E+12 & 908 ($\pm$9)   & 5.21  & 31.04 & 30.06 & 0.079 \\
    0.195 & 5.00E-08 & 0.074 & 1.06 ($\pm$0.04) E+11 & 51 ($\pm$3)    & 5.22  & 30.99 & 0.49  & 0.371 \\
    0.195 & 5.00E-08 & 0.074 & 1.49 ($\pm$0.04) E+11 & 74 ($\pm$3)    & 5.22  & 30.99 & 0.69  & 0.335 \\
    0.195 & 5.00E-08 & 0.074 & 2.14 ($\pm$0.04) E+11 & 115 ($\pm$2)   & 5.22  & 30.99 & 0.99  & 0.318 \\
    0.195 & 5.01E-08 & 0.074 & 3.29 ($\pm$0.04) E+11 & 169 ($\pm$2)   & 5.22  & 30.94 & 1.52  & 0.307 \\
    0.195 & 5.00E-08 & 0.074 & 5.05 ($\pm$0.04) E+11 & 227 ($\pm$2)   & 5.21  & 31.04 & 2.33  & 0.319 \\
    0.195 & 5.01E-08 & 0.074 & 6.37 ($\pm$0.05) E+11 & 273 ($\pm$2)   & 5.22  & 30.95 & 2.95  & 0.319 \\
    0.195 & 5.00E-08 & 0.074 & 8.79 ($\pm$0.06) E+11 & 331 ($\pm$2)   & 5.22  & 31.00 & 4.05  & 0.315 \\
    0.195 & 4.99E-08 & 0.074 & 1.11 ($\pm$0.01) E+12 & 371 ($\pm$3)   & 5.21  & 31.05 & 5.09  & 0.316 \\
    0.195 & 5.00E-08 & 0.074 & 1.42 ($\pm$0.01) E+12 & 418 ($\pm$3)   & 5.21  & 31.05 & 6.54  & 0.307 \\
    0.195 & 5.00E-08 & 0.074 & 2.23 ($\pm$0.02) E+12 & 533 ($\pm$4)   & 5.21  & 31.02 & 10.27 & 0.285 \\
    0.195 & 5.03E-08 & 0.073 & 4.22 ($\pm$0.02) E+12 & 696 ($\pm$4)   & 5.23  & 30.86 & 19.59 & 0.247 \\
    0.195 & 5.00E-08 & 0.074 & 6.93 ($\pm$0.05) E+12 & 856 ($\pm$6)   & 5.21  & 31.00 & 31.95 & 0.212 \\
    0.195 & 4.85E-08 & 0.074 & 8.40 ($\pm$0.08) E+12 & 904 ($\pm$8)   & 5.03  & 32.48 & 36.22 & 0.195 \\
    0.097 & 5.01E-08 & 0.005 & 3.47 ($\pm$0.06) E+12 & 627 ($\pm$12)   & 5.22  & 30.98 & 2.01  & 0.255 \\
    0.097 & 4.99E-08 & 0.005 & 8.66 ($\pm$0.05) E+12 & 871 ($\pm$6)   & 5.22  & 30.98 & 5.01  & 0.222 \\
    0.097 & 5.00E-08 & 0.005 & 1.73 ($\pm$0.01) E+13 & 1165 ($\pm$8)  & 5.22  & 30.98 & 10.00 & 0.158 \\
    0.097 & 4.98E-08 & 0.005 & 3.41 ($\pm$0.03) E+13 & 1486 ($\pm$12)  & 5.18  & 31.28 & 19.43 & 0.122 \\
    0.097 & 1.00E-08 & 0.115 & 2.26 ($\pm$0.03) E+12 & 122 ($\pm$3)   & 5.22  & 30.98 & 1.31  & 0.259 \\
    0.097 & 1.00E-08 & 0.115 & 3.47 ($\pm$0.03) E+12 & 218 ($\pm$5)   & 5.22  & 30.98 & 2.01  & 0.290 \\
    0.097 & 1.00E-08 & 0.115 & 5.20 ($\pm$0.07) E+12 & 297 ($\pm$8)   & 5.22  & 30.98 & 3.01  & 0.301 \\
    0.097 & 1.00E-08 & 0.115 & 7.80 ($\pm$0.12) E+12 & 421 ($\pm$5)   & 5.22  & 30.98 & 4.51  & 0.302 \\
    0.097 & 1.00E-08 & 0.115 & 8.66 ($\pm$0.09) E+12 & 463 ($\pm$4)   & 5.22  & 30.98 & 5.00  & 0.304 \\
    0.097 & 9.98E-09 & 0.115 & 1.16 ($\pm$0.03) E+13 & 546 ($\pm$5)   & 5.20  & 31.15 & 6.67  & 0.303 \\
    0.097 & 1.00E-08 & 0.115 & 1.73 ($\pm$0.07) E+13 & 715 ($\pm$10)   & 5.22  & 30.99 & 10.01 & 0.298 \\
    0.097 & 1.00E-08 & 0.115 & 3.46 ($\pm$0.02) E+13 & 1059 ($\pm$11)  & 5.22  & 30.99 & 20.00 & 0.265 \\
    0.097 & 1.00E-08 & 0.115 & 5.19 ($\pm$0.03) E+13 & 1316 ($\pm$15)  & 5.22  & 31.00 & 29.99 & 0.231 \\
    0.097 & 9.56E-09 & 0.115 & 6.80 ($\pm$0.05) E+13 & 1457 ($\pm$5)  & 4.95  & 33.20 & 35.60 & 0.224 \\
 \label{T:expt}
\end{longtable*}
\endgroup

\section{Error analysis\label{ch:app}}
The results presented in this paper are based on statistics of long time series of temperature measurements. A careful error analysis is warranted to extract appropriate error intervals for the mean values, which can then be propagated to errors in the dimensionless parameters.

The errors in the Rayleigh number, $Ra = \alpha_T g \Delta T H^3 / \nu \kappa$, and the Nusselt number, $Nu = q L / k \Delta T$, depend primarily on $\Delta T$, $T_\text{mean}$ and the error in heat flux, $q$. We formally propagate the errors for these quantities (e.g., Section 3.11 in \cite{Taylor97}), and report the uncertainties in $Ra$ and $Nu$ in Table~\ref{T:expt}. For the top and bottom temperatures, the spread in temperature across different thermistors is larger than that of any single thermistor, and so the error in $T_\text{bot}$ ($T_\text{top}$) is taken as the standard deviation $\sigma$ across the set of all bottom (top) thermistors. The error in $q$ is estimated as the standard deviation of the timeseries as well. Ultimately, the error bars in $Nu$ and $Ra$ are smaller than the marker sizes in the figures.

In contrast to the global parameters, interpretation of our laboratory $\midgrad$ results requires careful error analysis. For a series of independent observations $x$, the standard deviation of the mean $\sigma(\overline{x}) = \sigma / N^{1/2}$ is often used to characterize the error in $x$. In our study, though, temperatures are maintained through PID loop controls which enforce a roughly periodic behavior on time scales of $\mathcal{O}\left(10^3\right)$ seconds. To formulate proper error estimates we employ $N_\mathit{eff}$, the number of observations over which measurements are `effectively' independent \cite{Bayley46}. This replaces $N$ in the formulation for standard deviation of the mean \cite{Bayley46,Zieba11}:
\begin{equation}
\label{eq:stdm}
\sigma(\overline{x}) = \frac{\sigma}{N_\mathit{eff}^{1/2}} \, .
\end{equation}
\citet{Zieba11} write $N_\mathit{eff}$ in terms of elements of the autocorrelation function $r_k$, where:
\begin{equation}
\label{eq:neff}
N_\mathit{eff} = \frac{N}{1+2\sum\limits_{k=1}^{N_c}\left(1-\dfrac{k}{N}\right) r_k} \, .
\end{equation}
The summation is truncated at the so-called limiting lag, $N_c < N-1$. We determine the value of $N_c$ for each temperature timeseries by finding the first transit through zero of $r_k$ \cite{Zieba11}: 
\begin{equation}
\label{eq:ftz}
N_c = \text{min}\left\{k | \left(r_k>0~\wedge~r_{k+1}<0\right)\right\} \, .
\end{equation}
In each case, the error on each time-averaged thermistor measurement is given by Eq.~(\ref{eq:stdm}). Since temperature data are horizontally averaged, $\sigma(T)$ at every height is the root-mean-square combination of $\sigma$ for every thermistor at the same height. We then formally propagate errors in temperature into the normalized temperature profile following Eq.~(\ref{eq:grad}). Note that errors in not only the sidewall thermistors but also in $T_\text{bot}$ and $T_\text{top}$ figure into the error in $\overline{T}$.

$\midgrad$ is calculated as a best-fit slope for $\overline{T}$ vs. $z$, and so we employ the weighted least squares method to translate error bars in $\overline{T}$ into error bars in $\midgrad$ \cite{Taylor97}. For a set of sidewall heights $h_i$, corresponding to normalized temperatures $\overline{T}_i$ with standard deviations $\sigma_i$, the weight factors are given by $w_i = 1/\sigma_i^2$. The standard deviation in $\midgrad$ is then written as:
\begin{equation}
\label{eq:stdgrad}
\sigma_\text{gradient} = \left( \frac{\sum\limits_i w_i}{\sum\limits_i w_i \sum\limits_i w_i h_i^2 - \left(\sum\limits_i w_i h_i \right)^2} \right)^{1/2} \, ,
\end{equation}
which we interpret as the error bars for $\midgrad$. Errors are evidently small enough that trends in $\midgrad$ can be meaningfully discerned, though they may grow large as $\Delta T$ becomes small.

\bibliography{Bib_list}

\end{document}